\documentclass[aps,prx,amsmath,twocolumn,amssymb,nofootinbib,superscriptaddress,showpacs,notitlepage]{revtex4-2}

\usepackage[T1]{fontenc}
\usepackage[utf8]{inputenc}
\usepackage{color}
\usepackage{graphicx,epstopdf}
\usepackage{url}
\usepackage[normalem]{ulem}
\usepackage{amsthm}
\usepackage{amsmath}
\usepackage{epigraph}

\usepackage[dvipsnames]{xcolor}

\def\be{\begin{equation}}
\def\ee{\end{equation}}
\def\bea{\begin{eqnarray}}
\def\eea{\end{eqnarray}}
\def\bi{\begin{itemize}}
\def\ei{\end{itemize}}
\def\bin{\begin{enumerate}}
\def\ein{\end{enumerate}}

\begin{document}

\title{Quantum Random Number Generators : Benchmarking and Challenges}

%%%%%%%%%%%%%%%%%%%%%%%%%%%%%%%%%%%%%%%%%%%%%%%%%%%%%%%%%%%%%%%%%%%%%%%%%%%%%%%
\author{David Cirauqui}
\affiliation{Quside Technologies SL, Mediterranean Technology Park, 08860 Castelldefels (Barcelona), Spain}
\affiliation{ICFO - Institut de Ciencies Fot\`oniques, The Barcelona Institute of Science and Technology, 08860 Castelldefels (Barcelona), Spain}
\author{Miguel \'Angel Garc\'ia-March}
\affiliation{ICFO - Institut de Ciencies Fot\`oniques, The Barcelona Institute of Science and Technology, 08860 Castelldefels (Barcelona), Spain}
\affiliation{Instituto Universitario de Matem\'atica Pura y Aplicada, Universitat Polit\`ecnica de
Val\`encia, 46022 Val\`encia, Spain}
\author{Guillem Guig\'o Corominas}
\affiliation{ICFO - Institut de Ciencies Fot\`oniques, The Barcelona Institute of Science and Technology, 08860 Castelldefels (Barcelona), Spain}
%\author{Carlos Abellan}
%\affiliation{Quside Technologies SL, Mediterranean Technology Park, 08860 Castelldefels (Barcelona), Spain}
\author{Tobias Gra\ss}
\affiliation{ICFO - Institut de Ciencies Fot\`oniques, The Barcelona Institute of Science and Technology, 08860 Castelldefels (Barcelona), Spain}
\author{Przemys{\l}aw R. Grzybowski}
\affiliation{Institute of Spintronics and Quantum Information, Faculty of
Physics, Adam Mickiewicz University in Poznań, Umultowska 85, 61-614 Poznań, Poland}
\author{Gorka Mu$\tilde{\rm n}$oz-Gil}
\affiliation{ICFO - Institut de Ciencies Fot\`oniques, The Barcelona Institute of Science and Technology, 08860 Castelldefels (Barcelona), Spain}
\affiliation{Institute for Theoretical Physics, University of Innsbruck, Technikerstr. 21a, A-6020 Innsbruck, Austria}
\author{J. R. M. Saavedra}
\affiliation{Quside Technologies SL, Mediterranean Technology Park, 08860 Castelldefels (Barcelona), Spain}
\author{Maciej Lewenstein}
\affiliation{ICFO - Institut de Ciencies Fot\`oniques, The Barcelona Institute of Science and Technology, 08860 Castelldefels (Barcelona), Spain}
%\affiliation{ICREA - Instituci\'o Catalana de Recerca i Estudis Avan\c cats, E-08010 Barcelona, Spain}
\affiliation{ICREA, Pg. Llu\'is Companys 23, 08010 Barcelona, Spain}

\date{\today}

\begin{abstract}
We discuss the current state of the art of Quantum Random Number Generators (QRNG) and their possible applications in the search for quantum advantages. To this aim, we first discuss a possible way of benchmarking QRNG by applying them to the computation of complicated and hard to realize classical simulations, such as critical dynamics in two-dimensional Ising lattices. These are performed with the help of computing devices based on field-programmable gate arrays (FPGAs) or graphic processing units (GPUs). The results obtained for QRNG are compared with those obtained by classical pseudo-random number generators (PRNG) of various qualities. 
Monte Carlo simulations of critical dynamics in moderate lattice sizes  (128$\times$128) start to be sensitive to the correlations present in pseudo-random numbers sequences, allowing us to detect them. By comparing our analysis with that of Ref. [PRE {\bf 93}, 022113 (2016)], we estimate the requirements for QRNGs in terms of speed, rapidity of access, and efficiency to achieve the objective of quantum advantage with respect to the best PRNGs. We discuss the technical challenges associated with this objective.
\end{abstract}
\maketitle

\epigraph{\it We dedicate this work to the memory of Roy J. Glauber and Fritz Haake, once the Masters of Kinetic Ising Models}

%%%%%%%%%%%%%%%%%%%%%%%%%%%%%%%%%%%%
\section{Introduction}
%%%%%%%%%%%%%%%%%%%%%%%%%%%%%%%%%%%%

\noindent{\it Quantum Technologies.} The second and the third decade of the XXI century have witnessed the rapid developments of Quantum Technologies. In most countries, these developments are organized in four so-called "vertical pillars": quantum computation, quantum simulation, quantum metrology/sensing, and quantum communications \cite{Acin2018}. There is also a horizontal bar connecting the four pillars, focused on basic science for Quantum Technologies,  quantum software, fundamentals of quantum information science, or quantum effects in thermodynamic processes.

The universal fault-tolerant quantum computing with sufficient number of qubits and error corrections remains a dream and a severe fundamental and technological challenge. Nevertheless, the spectacular progress in noisy intermediate-scale quantum (NISQ) technologies allowed us to approach or even reach the quantum advantage (frequently termed quantum supremacy). Recently, the famous experiment by Google demonstrated efficient sampling from a random quantum circuit of 53 qubits \cite{Arute2019}. Similar results were achieved in photon sampling \cite{JWPan}. Quantum simulators have achieved quantum advantage already some time ago, mainly in studying problems that belong to the area of physics (cf. \cite{ATrab,LSA2017,LNP1000}), such as static and dynamical properties of quantum many-body systems (cf. \cite{Trotzky2012}). The outstanding successes of quantum metrology and sensing range from the use of squeezed states in gravitational wave detection \cite{Aasi13} to unprecedented precision in magnetometry (for a review, see \cite{Braun2018}). Quantum communications reached perhaps the highest technology level, as quantum cryptographic systems are widely used outside of academia and are commercially available \cite{Pirandola2019, Gisin2002}.

\noindent{\it Quantum Random Number Generators.} Since some applications of random numbers have to do with the security of information processing, Quantum Random Number Generators (QRNG), as a novel technology, belong to the pillar of quantum communication. Also, most of the available QRNG in academia or on the market use experimental methods and technological tools similar to quantum cryptography. The reasons for this significant interest and demand in QRNGs are manifold, including:
\begin{itemize}
\item All of the {\bf classical RNGs are pseudo-random}. Typically, there is a trade-off between the statistical correlations between pseudo-random numbers \textit{vs} their generation rate and efficiency. RNGs using iterative non-linear maps may be very fast but have typically relatively short correlation lengths. RNGs based on measuring physical or natural processes are correlated on much larger scales but are slower and less efficient. Nonetheless, combining iterative RNG with FPGA  or GPU technologies, as for instance discussed in \cite{Lin2016}, might circumvent these problems.

\item If quantum mechanics is correct (which we so far believe based on striking experimental evidence), then its predictions are {\it intrinsically random}.
In particular,  this randomness can be certified in a Bell test (which, however, itself requires the use of random numbers) \cite{Rowe2001}.
The philosophical, physical, and technological consequences of the randomness of quantum mechanics are discussed in great detail and from various points of view in the recent review by some of us \cite{Bera2017}. QRNGs thus distinguish themselves from a fundamental perspective as a {\bf remedy against pseudo-randomness}.

\item Recent technical progress allows for the construction of  {\bf faster and more efficient QRNGs}, thus allowing, for instance, the first loophole-free observations of the violation of Bell inequalities and the test of non-locality of quantum mechanics \cite{Hensen2015, Giustina2015, Shalm2015}.

\item As such, QRNGs promise numerous {\bf applications} in quantum cryptography in particular, as well as on areas going far {\bf beyond quantum communications}.

\end{itemize}

Indeed, within all the different areas in which Quantum Technologies could be innovative, Random Number Generation appears as one of the fields that may benefit most from applying these new technologies.

We repeat: contrary to traditional entropy sources (e.g., thermal sources), in which we generate entropy by exploiting our lack of knowledge of the internal state of the system, quantum entropy sources are random by principle. That means that, even if we knew the whole state of the system, the outcome of the measurements remains unpredictable due to the inherently probabilistic nature of Quantum Mechanics. Moreover, this probabilistic character is fundamental; out of principle, this randomness has no internal structure, making these entropy sources fundamental for deploying the ultimate random number generators.

These ultimate randomness sources are critically important, for example, when employing randomized algorithms: procedures for efficient solving specific problems that employ random decisions in some parts of the process. Since these procedures rely on randomness, employing low-quality entropy sources can severely degrade the algorithm's efficiency and its outcome.

\noindent{\it Plan of the paper}. This paper is organized as follows. In Section II, we shortly review the technological aspects of quantum randomness following the lines of \cite{Bera2017}. Section III explains how to benchmark RNGs using hard-to-compute, complex Monte Carlo simulations. To this aim, we devote various subsections to discuss the kinetic Ising models and the problem of calculating the dynamical critical exponent $z$. Last, we review the results of the Ref. \cite{Lin2016}, which stimulated our studies. Section IV contains the application of the latter methodology to PRNGs, whereas Section V presents the same methodology applied to a QRNG. Finally, Section VI presents a brief review of the literature regarding the calculation of the critical exponent; finally, we conclude in Section VII, presenting general requirements for the speed and efficiency of QRNGs to reach the quantum advantage.

\section{Quantum Randomness and technology}

Usually, when talking about the importance of random numbers, one talks about gambling and evokes the famous words of Julius Caesar: {\it "Alea iacta est"}. Still, {\it "the importance of random numbers in politics, social science and medicine should also not be underestimated; randomized polling and randomized trials are essential methodologies in these areas"} \cite{Bera2017}. One could even add here arts: randomness plays an important role in contemporary music \cite{serialism1, serialism2, aleatoricism1, aleatoricism2, yamada2021applications, sonar} and visual arts \cite{abstract}.

A significant challenge for randomness technologies consists of assuring non-predictability of RNG outputs - this is usually done with statistical tests \cite{rukhin2001,marsaglia2002, eddelbuettel2007}, which, however, have serious drawbacks. One can summarize these drawbacks in the famous cartoon joke of Calvin \footnote{In the cartoon, somebody presents Calvin as a new RNG. Calvin says {\it Nine, nine, nine, nine, nine...} The question is: is it truly random? You never know with randomness.}; or in the more profound saying of John von Neumann, {\it there is no such thing as a random number -- there are only methods to produce random numbers} (von Neumann, 1951).

In view of these profound problems, and {\it "in light of the difficulties in determining the predictability of the apparent randomness seen in thermal fluctuations and other classical phenomena, using the intrinsic randomness of quantum processes is very attractive"} \cite{Bera2017}. 

Recently, several methods have been proposed as QRNG candidates:
\begin{itemize}
\item Using device-independent (DIQIP) randomness protocols employing  Bell-inequality violation, using for instance  ions~\cite{Pironio2010}, photons \cite{Giustina2015, Shalm2015}, nitrogen-vacancy centres \cite{Hensen2015},
neutral atoms \cite{Rosenfeld2009} and superconducting
qubits \cite{Jerger2016}. These lead to the so-called "certified randomness" \cite{Acin2017}.

\item Nevertheless, DIQIP approaches are not very efficient and fast. For practical reasons, one can use signals from quantum processes, and devices to harness the intrinsic randomness of quantum mechanics -- these have existed since the 1950s. These involve devices to observe the timing of nuclear decay, \cite{Isida1956} electron shot noise in semiconductors, splitting of photons on beamsplitters, timing of photon arrivals, vacuum fluctuations, laser phase diffusion, amplified spontaneous emission, Raman scattering, atomic spin diffusion, and others. For a review on the topic, see Ref.~\cite{Herrero2017}.

\item As for today, the fastest quantum random number generators are based on laser phase diffusion \cite{Abellan2014, Jofre2011, Xu2012, Yuan2014}, with the record at the time of writing being 68~Gbps \cite{Nie2015}. These devices, illustrated in Fig. 1, {\it "work entirely with macroscopic optical signals (the output of lasers), which greatly enhances their speed and signal-to-noise ratios. It is perhaps surprising that intrinsic randomness can be observed in the macroscopic regime, but in fact, laser phase diffusion (and before it maser phase diffusion) was one of the first predicted quantum-optical signals, described by Schawlow and Townes in 1958 (Schawlow and Townes, 1958)"}. \cite{Bera2017}.
\end{itemize}

\begin{figure}
 \includegraphics[width=\columnwidth]{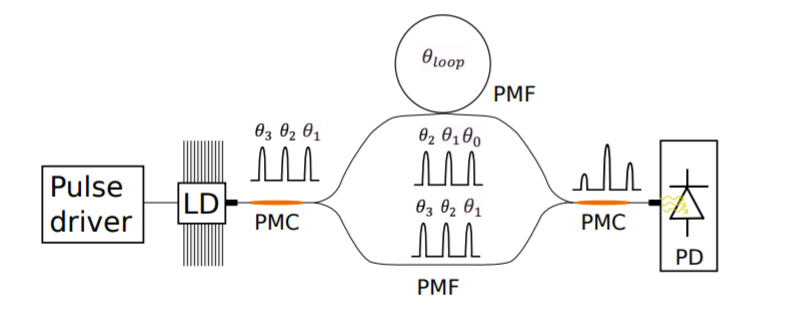}
 \caption{{\bf Scheme of Quside's PD-QRNG used in this work.} A laser is modulated above and below threshold, each time generating a pulse with a random phase $\theta_i$. By means of an unbalanced Mach-Zehnder interferometer, each pulse is interfered with a later generated random phase pulse, turning phase fluctuations into intensity fluctuations, which are further converted into random numbers by using conventional photodetectors and electronics. Image taken with permission from \cite{Abellan2014}.}
  \label{fig:fig1}
\end{figure}

A possible way to characterize the RNG device is via the conditional min-entropy \cite{Mitchell2015},
\begin{align}
 H_{\infty}&(X_i|h_i)\geq k,\,\,\,\,
\forall i \in \mathbb{N},\,\, \forall h_i,
\end{align}
which gives a lower bound to the unpredictability of a set of outcomes $X_i$ from the RNG devices , given the device's history $h_i$, at that moment (which includes all fluctuating quantities not ascribable to intrinsic randomness). If the conditional min-entropy is bounded from below, randomness extraction techniques can be used to produce arbitrarily-high-quality output bits from the considered source.

In fact, determining the min-entropy due to intrinsic randomness of laser phase-diffusion QRNGs was a very challenging task \cite{Mitchell2015}, especially in the context of Bell tests \cite{Abellan2015}. Thanks to these efforts and the modeling and measurement considerations, it was possible to bound the min-entropy of these devices. In effect, laser phase diffusion random number generators have been used successfully in all loophole-free Bell tests \cite{Giustina2015,Hensen2015,Shalm2015}. % Here we outline of the output of these devices.

%\MAGM{Remember to put a comment that links to next section}

%\noindent{\bf Earlier results} In Ref. \cite{Lin2016} the authors study the critical dynamics in two-dimension Ising lattices up to $L = 2048$ lattices using state-of-art field programmable gate arrays (FPGA).  They measure the linear relaxation times from extremely long Monte Carlo simulations. The longest simulation has 7.1 $\times 10^{16}$ spin updates, which, according to the authors, would take over 37 years to simulate on a general purpose computer. The linear relaxation time of the Ising lattices is found to follow the dynamic scaling law for correlation lengths as long as 2048. The dynamic exponent $z$ is found to be $2.179(12)$, in agreement with previous works on  Ising lattices \cite{linke1995}. From the perspective of the present paper of ours the most important observation is that Monte Carlo simulations of critical dynamics in Ising lattices of moderate sizes ($L=128)$ start to be very sensitive to the statistical correlations between pseudo-random numbers, making it even more difficult to study such large systems. The main objective of the present study is indeed to compare as fairly as possible  the results of Ref. \cite{Lin2016} to ours, obtained using QRNGs.

\section{Methodology}
\label{sec:meth}

In this work, we propose to benchmark RNGs or QRNG by using large sequences of random numbers to perform challenging calculations, such as the Monte Carlo simulation of a complex system, whose result may be largely affected by the quality of these numbers. One should then: i) calculate a physical quantity particularly sensitive to the statistical properties of random numbers used; ii) check the results with respect to convergence and self-consistency; iii) compare the results for different RNGs, determining in this way "the best ones". This approach has been recently successfully used by Lin and Wang \cite{Lin2016}, who applied it to the calculations of the dynamical critical exponent $z$ for a two dimensional kinetic Ising model. Below we describe with necessary details kinetic Ising models, dynamical scaling, Monte Carlo implementations and, finally, the results of Ref. \cite{Lin2016}.\\

\noindent{\bf Physical problem: kinetic Ising model}.
\begin{figure*}
 \includegraphics[width=1.8 \columnwidth]{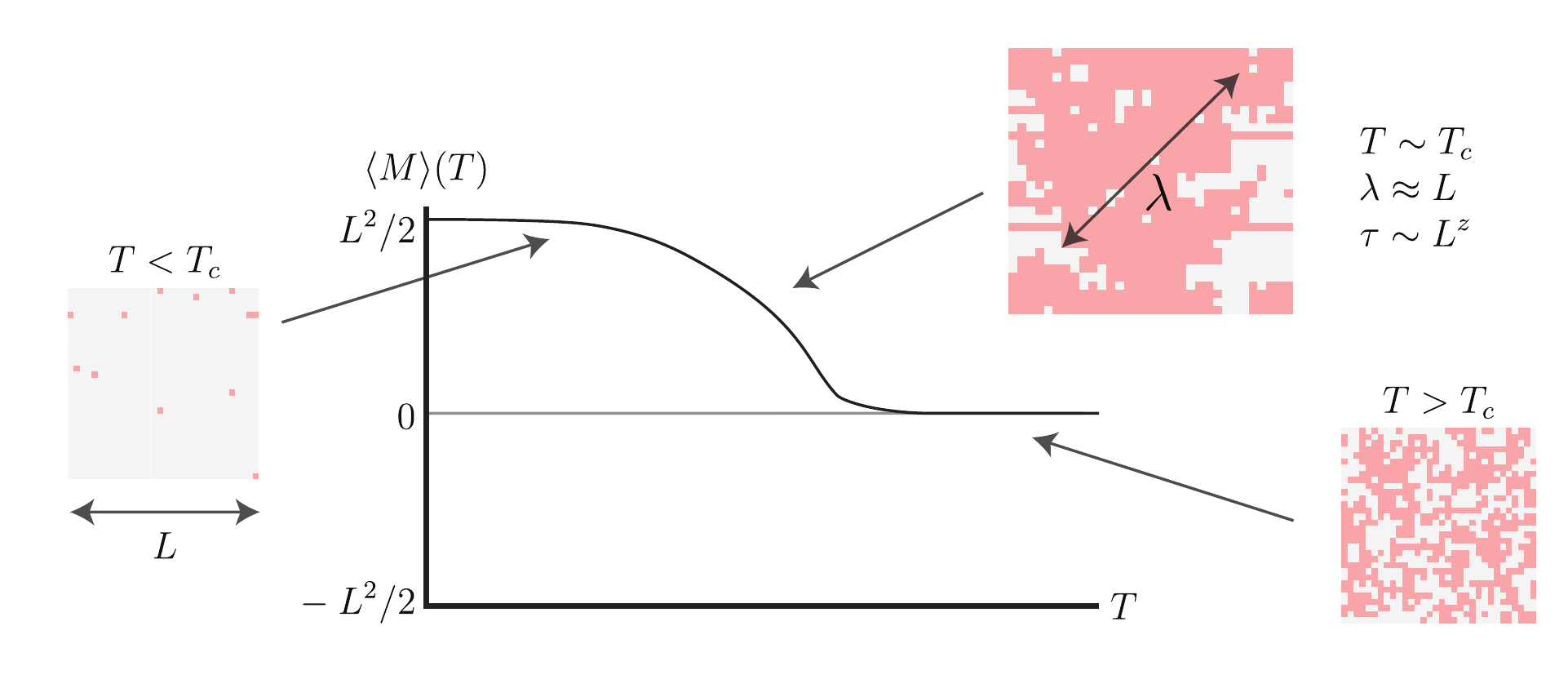}
 \caption{{\bf 2D Ising model dynamics.} For temperatures below the critical temperature, $T_c$, the system shows ferromagnetic behavior; for temperatures larger than $T_c$, the system shows ferromagnetic behavior. At $T_c$, the system shows a phase transition (see main text for details).}
  \label{fig:fig2}
\end{figure*}
Randomized algorithms, such as Monte Carlo codes, are applied nowadays to a plethora of problems, ranging from the practical problems of relevance for  industry and society to the purely academic ones.
Among the latter, the calculation of the critical exponents associated with the phase transition of a 2D Ising model is noteworthy.
In this system, there is a set of $N$ Ising spins ${\sigma_i}=\pm 1$ arranged on a two-dimensional square lattice of linear side $L$, such that $N=L^2$. Spins are coupled  by means of the coupling coefficient $J_{ij}$. We consider only coupling  to nearest neighbors and $J_{ij}=J$ constant for all pairs of spins. The Hamiltonian that describes the statics and thermodynamics  of this system is therefore
\begin{equation}
\label{eq:hamiltonian}
H = -\sum_{i,j} J_{ij} \sigma_i \sigma_j = - J \sum_{\left\langle i,j\right\rangle}  \sigma_i  \sigma_j,
\end{equation}
where the sum of the term on the right is restricted to the nearest neighbours, and $1 \leq i,j \leq L$.
This model is commonly used to study the behavior of ferromagnetic materials; for this purpose, periodic boundary conditions are typically used, simulating a square unit cell of size $L$ within the periodic system under study.

The model describes the transition between the two phases of a simple ferromagnetic material: at low temperatures, most of the spins are aligned in the same direction, resulting in non-zero magnetization and ferromagnetic behavior of the material at macroscopic scales; on the contrary, at high temperatures, the fluctuations of the spins associated to thermal effects exceed the ferromagnetic order induced by the Hamiltonian; in this scenario, the spins act independently, generating zero average magnetization and thus making the material paramagnetic.
The phase transition occurs when the system reaches the so-called critical temperature $T_c$; at this temperature, the domains of high magnetization in one direction are gradually destroyed by the effect of thermodynamic fluctuations, and converted into zones of high magnetization in the opposite direction. Then, macroscopically, the system is magnetized, but the orientation of the magnetic field changes in time. %we call the characteristic fluctuation time -- the coherence time, $\tau_k$. 
The time-delayed correlation of the order parameter (magnetization) is
\begin{equation}
\chi(t)=\langle M(t)M(0)\rangle
=\sum_k a_k e^{-t/\tau_k},
\end{equation}
where the sum over $k$ runs over the system's excited modes~\cite{Glauber1963}, $M$ is the system's magnetization, $\tau_k = 1/\lambda_k$ is the relaxation time
for the $k$-th excited mode, which has eigenvalue $\lambda_k$, and $a_k$  are $t$-independent constants~\footnote{The probability of finding the system at configuration $\sigma$ at time $t$ for some initial configuration being $\sigma_0$ can be solved to be $P(\sigma,t|\sigma_0)=\sum_{k=0}^\infty c_k(\sigma_0)e^{-\lambda_k t}\phi_k(\sigma)$ with $\phi_k(\sigma)$ being the eigenmode with eigenvalue $\lambda_k$ with standard methods (see \cite{Glauber1963} for one dimension and via Master equation, and more general methods in \cite{ma2018,zinn1996})}.

If the system is truly  infinite,  at the temperature approaching the
critical temperature $T_c$, the relaxation times $\tau_k$ diverge leading to  the "critical
slowing down" effect. The dynamic critical
scaling hypothesis \cite{ma2018, zinn1996, halperin1969, hohenberg1977,suzuki1977,2004Odor}  predicts that the diverging $\tau_k$ have a power-law dependence on the diverging static correlation length
$\xi$ which scales as  $\xi \propto  |T - T_c|^{-\nu}$. We define then the  dynamic exponent $z$ as
\begin{equation}
\tau_k \propto  \xi^z.
 \label{eq:tau}
\end{equation}
Since all $\tau_k$ diverge, the relaxation process is dominated by the first excited mode with $\tau_1$. This one effectively determines  the relaxation time of the system $\tau=\tau_1$ of the whole system, so that
\begin{equation}
\chi(t)=\langle M(t)M(0)\rangle \propto e^{-t/\tau}.
\label{eq:chi}
\end{equation}

In practice, we deal with finite systems of size $L^2\propto M^2 =N$. This means that at $T_c$, the static correlation  $\xi$ grows achieving the longest wavelength $\lambda_{\rm max} \simeq L$. The relaxation time of the system is then expected  to have a power-law relation with the $\lambda_{\rm max}$, i.e. for finite size scaling (FSS)
\begin{equation}
\tau_{\rm{FSS}}(L) \propto L^z.
 \label{eq:tauFSS}
\end{equation}
Combining the above equations, one can determine the relaxation times and the exponent $z$ of finite Ising lattices from Monte Carlo simulations of kinetic models, as first suggested by Hohenberg
\cite{hohenberg1977}.

\noindent{\bf Benchmarking of randomness.} 
The Kinetic Ising model~\cite{Glauber1963} presented above  belongs to a universality class called model A, or Ginzburg-Landau stochastic models without energy conservation, which are the simplest models showing slowing down critical dynamics with no conservation laws. As previously shown, this slowing down is characterized by the critical exponent $z$~\cite{hohenberg1977}. Its relaxation dynamics can be described in various ways: i) by a master equation for the time evolution of the probability distribution for one of the possible configurations of the system of spins at time $t$; ii) by Markovian stochastic dynamics fulfilling Fluctuation-dissipation $t$~\cite{Glauber1963}; iii) by a Fokker-Plank equation~\cite{ma2018,zinn1996}. In this sense, a large body of research has pursued finding approximations for $z$ via a variety of field theory approaches~\cite{1972HalperinPRL,1975DeDominicisPRB,1976RaczPRB,1981BauschPRL,1984DomanyPRL,1993DammannEPL,1993WangPRB,1998PrudnikovPSS,1998WangPRE,2006FolkJPA,2006KrinitsynTMP,2007CanetJPA,2009NalimovTMP,2012LubetzkyCMP,2015MesterhazyPRD,2017DuclutPRE,2022AdzhemyanPLA,2021SilvanoArxiv}. To the best of our knowledge, the most recent outcome from this approach is $2.14(2)$ for the model in two dimensions~\cite{2022AdzhemyanPLA}.  
The determination of $z$ can also be achieved via Monte Carlo (MC) simulation of the dynamics of a suitable kinetic two-dimensional Ising model, as discussed above. Many works have pursued, for more than forty years, the accurate estimation of $z$~\cite{1976BoltonPRB, 1985WilliamsJPA,1987ItoJPSJ,1987TangPRB,1988ItoJPC,1988MoriPRB,1991FerrenbergJSP,1992StaufferPhysA,1993ItoPhysA,1993MunkelPhysA,1995GrassbergerPhysA,1995GropengiesserPhysA,1995LiPRL,1996LiPRE,Nightingale1996,1997SoaresPRB,1997WangPRE,2000GodrecheJPA,2000ItoJPSJ,2000NightingalePRB,2007LeiCSB}. 
 Due to the difficulty of these problem, relying on huge numbers of \textit{random}, MC steps, the estimation of $z$ is highly sensitive to the quality of the random numbers employed for the simulation. The presence of correlations in random numbers may combine with the correlations generated by the kinetic Ising dynamics and the Hamiltonian themselves, resulting in mean values and variances of the critical exponent that deviate from the real value. The latter will then heavily depend on the quality of the RNG. These effects can be benchmarked by comparing the resulting $z$ with those that would arise using a perfect randomness source.

\noindent{\bf  Metropolis algorithm.}
Kinetic Ising Models have been studied for decades. Yet, there is no known analytical solution for two and higher dimensions: in these cases, one has to rely on approximate or numeric methods. In our study, we make use of the Metropolis algorithm \cite{Metropolis1953}, due to its simplicity and ease of generating an efficient implementation on hardware accelerators, such as field-programmable gate arrays (FPGAs) and Graphical processing units (GPUs). For the Metropolis algorithm, a given spin flips whenever its contribution to the total energy is reduced. However, it is also possible that the spin flips against energy minimization, subject to a certain probability that depends on the temperature. Schematically, the algorithm consists in:

\begin{itemize}
\item Propose a spin flip $\sigma_i \rightarrow -\sigma_i$ and compute the energy difference $\Delta E$. If $\Delta E \le 0$, accept the update.
\item Otherwise, toss a random number $0\le r\le 1$ and accept the update if $r\le \exp(-\beta \Delta E)$.
\end{itemize}
Note that it requires an extensive use of random numbers, on the order of $N^2$ for one MC  step per spin (this is, applying the previous recipe for all spins in the system). As we approach the critical point, the number of MC steps necessary to reach equilibrium goes to infinity due to the critical slowing down effect discussed above.

In order to parallelize computations, we split an $L$ by $L$ spin-lattice (with $L$ a power of two, for the sake of simplicity of the implementation) in two different colored sublattices, in a chessboard-like fashion. This way, for a given spin, all its nearest neighbors are of its opposite color (i.e., they lie in a different sublattice) \cite{Weigel2017}. This fact prevents equally-colored spins from interacting directly and therefore allows us to update all the elements of the same sublattice in parallel. Thus, each Monte Carlo sweep is realized in two steps, each of them updating one of the sublattices in parallel; following the chessboard analogy, updating first the black squares, then the white ones.

\noindent{\bf FPGA implementation}
The Metropolis algorithm can be implemented directly on hardware acceleration platforms, such as  field-programmable gate arrays (FPGAs). Several works have exploited FPGAs or, similarly, Graphical processing units (GPUs) to simulate spin systems (see. e.g. \cite{Lin2016,2007HerbordtC,2008BellettiCPC,2009PreisJCP,2010BlockCPC,Weigel2012,2013LinJCP}). By these means, we are able to directly integrate both the Ising model simulator and the quantum random number generator hardware on the same computing device, thus removing any performance bottleneck associated with data communication.

This generator, designed by the company Quside Technologies, uses the phase diffusion phenomenon associated with a laser source as its source of quantum randomness (see Fig.~\ref{fig:fig1}). In broad terms, when the laser system is off, due to the uncertainty principle, the phase is undetermined. When the laser switches on, one of the phases is selected at random, thus defining the global phase of the laser pulse. Through laser modulation, this cycle repeats over and over again, generating a new phase that is purely random and decorrelated to the phase of the previously generated pulse.

In order to recover the phase of each pulse (which is the quantum random variable), the pulses are made to interfere with their immediate preceding ones in an unbalanced Mach-Zehnder interferometer (uMZI). The combination of amplitudes hence allows converting phase fluctuations into intensity fluctuations, which can be easily detected by a conventional photodetector. These intensity fluctuations are subsequently digitized and converted into a stream of random numbers, which feeds the Ising model simulator, thus increasing the efficiency of the system.

\begin{table}
    \centering
    \begin{tabular} { |c|c|c|c| }
        \hline
        & Modulus & Multiplier & Increment\\
        \hline
        PRNG0 & $2^{32}-1$ & $16807$ & 0\\
        \hline
        PRNG1 & $2^{25}-39$ & $12836191$ & 0\\
        \hline
        PRNG2 & $2^{23}-15$ & $422527$ & 0\\
        \hline
        PRNG3 & $2^{17}-1$ & $43165$ & 0\\ [1ex]
        \hline
    \end{tabular}
    \caption{Parameters used in the Linear Congruential Generators  used.}
    \label{tab:tab1}
\end{table}

\section{ Results and discussion}

For the calculation of the critical coefficient, we simulate the spin model of Eq.~\ref{eq:hamiltonian} for a time of $1300\tau$ and calculate its magnetization after each MC step. After inspection of Eqs.~\eqref{eq:tau} and \eqref{eq:tauFSS}, we approximate $\tau$ here by $L^{z_{\rm{approx}}}$ with $z_{\rm{approx}}=2$. Note that $\xi$, the expected theoretical value of the correlation length,  equals $L$, since $\tau\propto\xi^z$ and $\tau_{\rm{FSS}}(L) \propto L^z$. From these results, the different values of the correlation time $\tau_{\rm{FSS}}(L)$ have been obtained by adjusting the correlation to a decreasing exponential for a time interval $t \in (0.3\tau_{\rm{FSS}}, 1.1\tau_{\rm{FSS}})$, in order to avoid both the initial high non-linearities and the fluctuations in the tail of the exponential, following Ref.~\cite{Lin2016}. 
\\
\\
\subsection{Detecting correlation effects in PRNGs}
%\section{Correlations in RNG}

As stated above, the dynamic exponent $z$ is known to be sensitive to the correlations of the random numbers employed in its computation. Therefore, in order to study the effects that such correlations have in our results, we make use of different PRNGs to simulate our system. 

\paragraph{ Fundamental tests with different PRNGs}

Before introducing the calculation with the QRNGs, we address the computation of the dynamic exponent with four different linear congruential generators (LCGs)-based generators, all of them presumably showing low correlations \cite{Lecuyer1999}. Generally, an LCG of the form $x_{i+1} = \left(ax_i + c\right) {\rm mod}\ m$ is described by three different parameters: its modulus $m$, its multiplier $a$, and increment $c$. The modulus sets its period of repetition; once selected, the other two parameters may be tuned such that the generator exhibits low correlations.

Regarding the simulation procedure, there are a couple of questions worth commenting. The first one regards an estimation of the total amount of random numbers required per lattice sweep, which in our case will be $n_{\rm{par}} = L^{2}$, for a lattice with $L$ sites per side (see Appendix~\ref{sec:numberofP} for a detailed discussion on the the quantity of random numbers required by our codes, and possible routes for future optimization). Second, given the suitability of GPU architectures for solving Ising-like problems, we implement our PRNGs and kinetic Ising algorithms in such a device in order to test the performance of the different PRNGs. We can also take advantage of its single-instruction multiple-thread (SIMT) execution model in order to avoid the pseudo-random number generation bottleneck. To do so, we implement a GPU kernel that, given the $L^{2}$ seeds, produces $L^{2}$ new PRNs in parallel, employing highly tested and consolidated LCGs that exhibit low correlations \cite{Lecuyer1999}. Using LCGs also allows us to tune their parameters easily, and therefore explore the effect that the correlations appearing in PRN sequences have on the results of our MC simulations.

With each PRNG and for each lattice size (ranging from $L = 4$ to $L = 512$), we compute the dynamic exponent $z$ with a Monte Carlo simulation. In Table~\ref{tab:tab1} we give the parameters used for each generator \cite{Lecuyer1999}.
 For each point in the simulation, we run a total number of $N_{it} \geq 100$ iterations in order to extract statistically relevant results (the number of iterations is restricted to $N_{it}\sim10$ for $L = 512$ lattices, due to long computation times). We fit the obtained curves with exponential laws in order to obtain $\tau$ for different lattice sizes, and then plot them in a logarithmic scale as a function of the lattice size to obtain the dynamic exponent $z$. We then compare them with the  theoretical estimate of $z$ obtained by studying the stochastic matrices governing the physics of our system in the classical Ref.~\cite{2000NightingalePRB}, that is, $z = 2.1667 \pm 0.0005$. We summarize our results in Table~\ref{tab:tab2}, in which we show the obtained dynamic exponents $z$, as well as their respective errors relative to the theoretical estimate, $\epsilon_r$  (we take as reference $z_{\rm ref}=2.1667$ and approximate  $\epsilon_r$  in the fourth decimal).

\begin{table}[ht]
    \centering
    \begin{tabular}{|c|c|c|}
         \hline
         & $z$ & $\epsilon_r$ \\
         \hline
         PRNG0 & $2.1087$ & $0.0268$\\
         \hline
         PRNG1 & $2.1159$ & $0.0234$\\
         \hline
         PRNG2 & $2.1047$ & $0.0286$\\
         \hline
         PRNG3 & $2.1162$ & $0.0233$\\
         \hline
    \end{tabular}
    \caption{Dynamic exponents $z$ and relative errors with respect to theoretical estimate $\epsilon_r$, obtained for each PRNG.}
    \label{tab:tab2}
\end{table}

All tested PRNG's yield a reasonable approximation of the dynamic exponent, as can be seen in Table~\ref{tab:tab2}, but they differ in their predicted value in the second significant decimal. 

\begin{figure}
    \centering
    \includegraphics[width=\columnwidth]{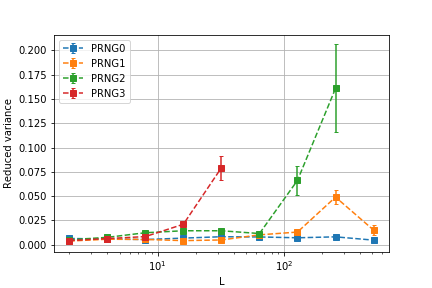}
    \caption{Statistical variance of the dynamic exponent $z$, normalized to its mean value, obtained from $N_{it} \geq 100$ iterations for all points except for $L=512$, for which $N_{it} = 30$, as a function of lattice size, for different PRNG's}
    \label{fig:fig3}
\end{figure}

Importantly, the statistical variance of the results poses a considerable caveat that must be taken into account when interpreting them. For any given PRNG, once the amount of random numbers consumed by the algorithm exceeds its period of repetition (modulus in Table~\ref{tab:tab1}), the sequence repeats itself, introducing a large amount of correlation at large time scales. We observe that these extra correlations affect the obtained dynamic exponent by exponentially enlarging the variance between iterations of the same simulation while keeping its mean value constant. To quantify these correlations, we define the normalized variance as $\sigma^2/\mu^2$, where $\sigma$ is the variance of the results and $\mu$ their mean value. For this case, the normalized variance stays more or less constant (around a value of $\frac{\sigma^2}{\mu^2} \approx 0.01$), until the lattice reaches a size large enough so that its simulation requires more random numbers than the generator's period of repetition. Beyond this point, the variance starts to increase exponentially with lattice size, as shown in Fig.~\ref{fig:fig3}. Furthermore, as we keep increasing $L$ far beyond the variance's explosion point, the obtained magnetization auto-correlation function $\chi\left(t\right)$ does no longer resemble an exponential decay, but instead starts showing a  noisy behavior (not shown). Therefore one cannot strictly speak of, nor extract, a reliable value for the dynamic exponent.

\begin{figure}
    \centering
    \includegraphics[width=\columnwidth]{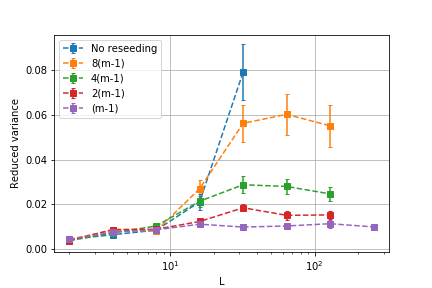}
    \caption{Normalized variance for PRNG3 as a function of lattice size. Physical reseeding of the generator every $\left(m-1\right)$, $2\left(m-1\right)$, $4\left(m-1\right)$, $8\left(m-1\right)$ random numbers. PRNG3 without reseeding is shown for comparison.}
    \label{fig:fig4}
\end{figure}

\paragraph{ Effects of reseeding.}
We now explore a natural question: how does the reseeding of the generators affect the results presented in previous paragraph, since the addition of physically random bits yields a theoretically infinite-period PRNG \cite{Lin2016}? We again study the behavior of the normalized variance in our Ising system by physically reseeding the same PRNG3 used in the previous section (which itself constitutes a computationally hard task). We reseed it every $\kappa\left(m-1\right)$ pseudo-random numbers, where $m$ is the modulus parameter of the LCG, and we first consider $\kappa = \left\{1, 2, 4, 8\right\}$. As shown in Fig.~\ref{fig:fig4}, we find that, for all cases, the variance explosion observed before is avoided. In this case, the normalized variance does not grow monotonically, but instead reaches a plateau whose value appears to be proportional to the number of repetitions introduced in the pseudo-random sequences before the reseeding is carried out (i.e., the amount of extra correlations introduced in our algorithm, compared to the infinite-period PRNG).

\begin{figure}[ht]
    \centering
    \includegraphics[width=\columnwidth]{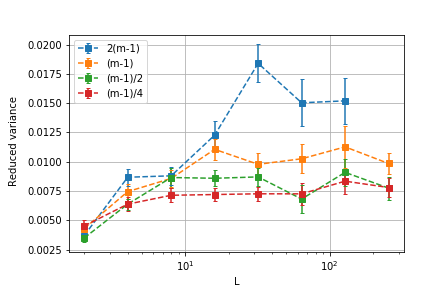}
    \caption{Normalized variance for PRNG3 as a function of lattice size. Physical reseeding of the generator every $2\left(m-1\right)$, $\left(m-1\right)$, $\frac{\left(m-1\right)}{2}$ and $\frac{\left(m-1\right)}{4}$ random numbers.}
    \label{fig:fig5}
\end{figure}

As shown in Fig.~\ref{fig:fig4}, the case $\kappa = 1$, i.e. an example of  an infinite-period pseudo-random number generator, exhibits the lowest normalized variance of all presented cases. Interestingly, by allowing $\kappa < 1$ (and thus paying the computational cost associated with a high-frequency reseeding), we observe that we can still lower this value further down (see Fig.~\ref{fig:fig5}). This scenario minimizes the correlations appearing in the pseudo-random sequences, approaching a true RNG (TRNG) as $\kappa$ gets smaller. Therefore, this fact allows us to conclude that {\it the use of PRNG's, even those showing low correlations and having a theoretically infinite period, can indeed affect the quality of our results in terms of variance} for the problem at hand. Moreover, as we show here, once $\kappa$ is fixed, the variance of the obtained $z$ converges to a plateau and barely changes with $L$. Hence, performing longer simulations will not improve the approximation. 

\begin{figure}[ht]
    \centering
    \includegraphics[width=\columnwidth]{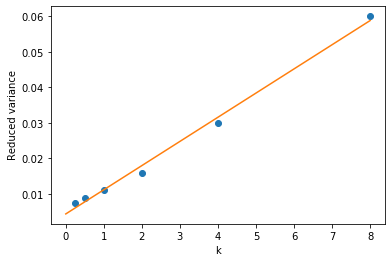}
    \caption{Reduced variance for different values of the reseeding period $\kappa$ of PRNG3 (dots) and linear fit (line).}
    \label{fig: fit var(k)}
\end{figure}

Next, we extract the values of the reduced variance plateaus and plot them as a function of the reseeding period $\kappa$. Interestingly, as shown in Fig.~\ref{fig: fit var(k)}, both quantities show a linear relation. By means of a linear fitting, we then can extrapolate the reduced variance for $\kappa=0$, this is, the case of a TRNG. We obtain a value of $\frac{\sigma^2}{\mu^2}(\kappa=0)\approx0.004$, which coincides with the initial points of the curves of Figs. \ref{fig:fig4} and \ref{fig:fig5}, corresponding to simulations of very small latices, where correlations between pseudo random numbers are still non-detectable.  

On top of the presented effect on the variance of the estimation of the dynamic exponent $z$, we also analyze its mean value for different $\kappa$. We observe that the reseeding of the generators yields a more accurate result for the mean value of the dynamic exponent (see Table~\ref{tab:tab3}). We note that, for every case with reseeding, and for every frequency $\kappa$ implemented, the relative error to the theoretical value is reduced to about half the one obtained with the different PRNG's used in Table~\ref{tab:tab2}, which were not reseeded. Nevertheless, we cannot reduce this error further. We emphasize that reducing $\kappa$ reduces the variance, bringing the result closer to those of a TRNG, but at a larger computational cost for smaller and smaller $\kappa$. The results here point out that in the limit of very small $\kappa$, one should obtain  results close to those obtained with a TRNG. Nonetheless, there is no practical way of testing this due to the large computational cost.

\begin{table}[ht]
    \centering
    \begin{tabular}{|c|c|c|}
        \hline
         & $z$ & $\epsilon_r$  \\
         \hline
         $\kappa = 2$ & $2.1815$ & $0.0068$\\
         \hline
        $\kappa = 1$ & 2.1477 & 0.0088 \\
        \hline
        $\kappa = \frac{1}{2}$ & 2.1482 & 0.0085 \\
        \hline
        $\kappa = \frac{1}{4}$ & 2.1441 & 0.0104 \\
        \hline
    \end{tabular}
    \caption{Dynamic exponents $z$ and relative errors with respect to theoretical estimate $\epsilon_r$, for different reseeding frequencies of PRNG3.}
    \label{tab:tab3}
\end{table}

%%%%%%%%%%%%%%%%%%%

\begin{figure*}[ht]
 \includegraphics[width=2 \columnwidth]{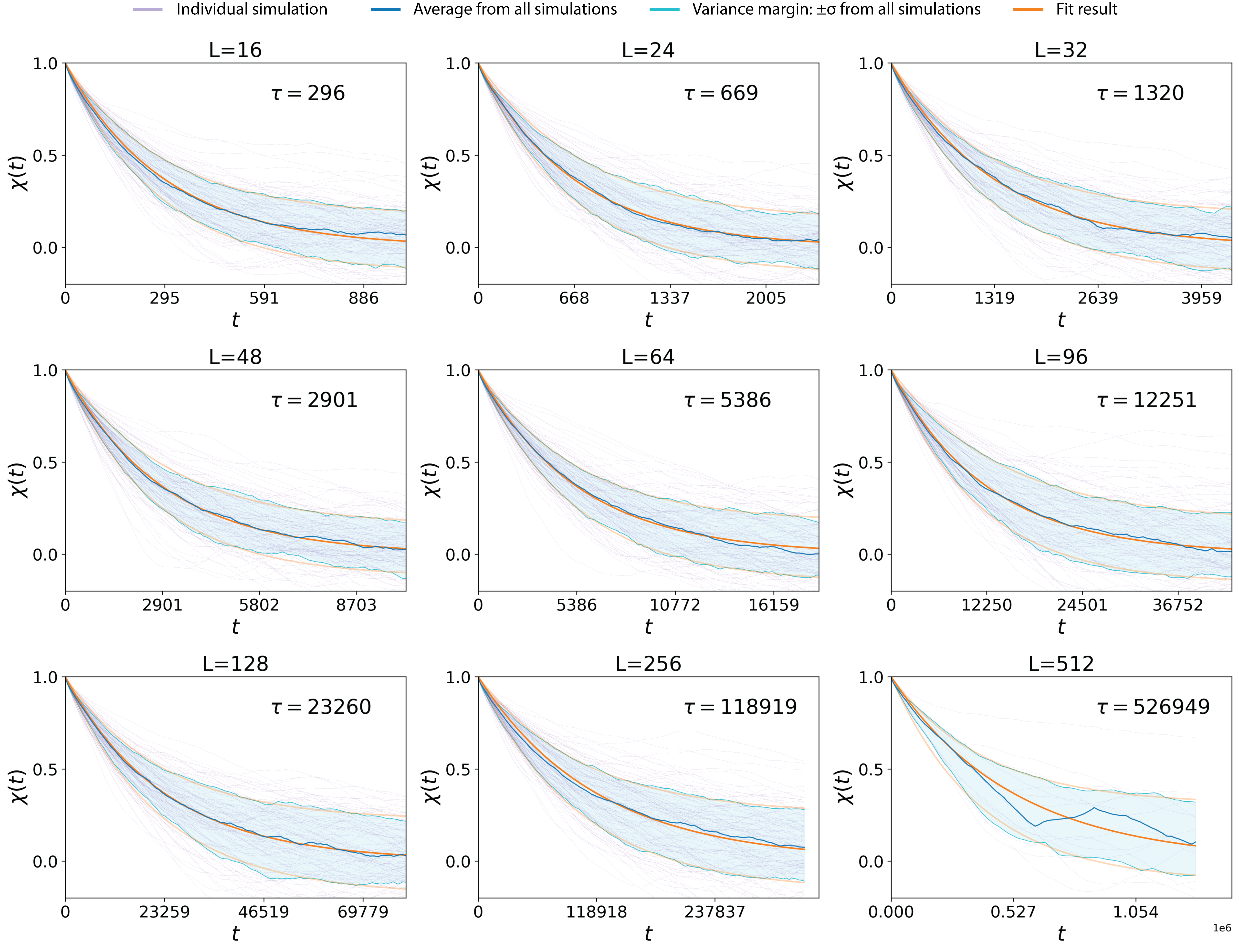}
 \caption{{\bf Determination of the relaxation time as a function of lattice size.} Inset $\tau$ values are the decay times associated to the orange curves, which correspond to the averages obtained by the fitting of the multiple repetitions simulated at each side length. The individual simulations (purple), as well as the variance interval for all of them (cyan) are also shown in the graphs.}
  \label{fig:fig7}
\end{figure*}

\section{Calculation of the dynamic critical exponent $z$ with a QRNG}
\begin{figure}[ht]
 \includegraphics[width=\columnwidth]{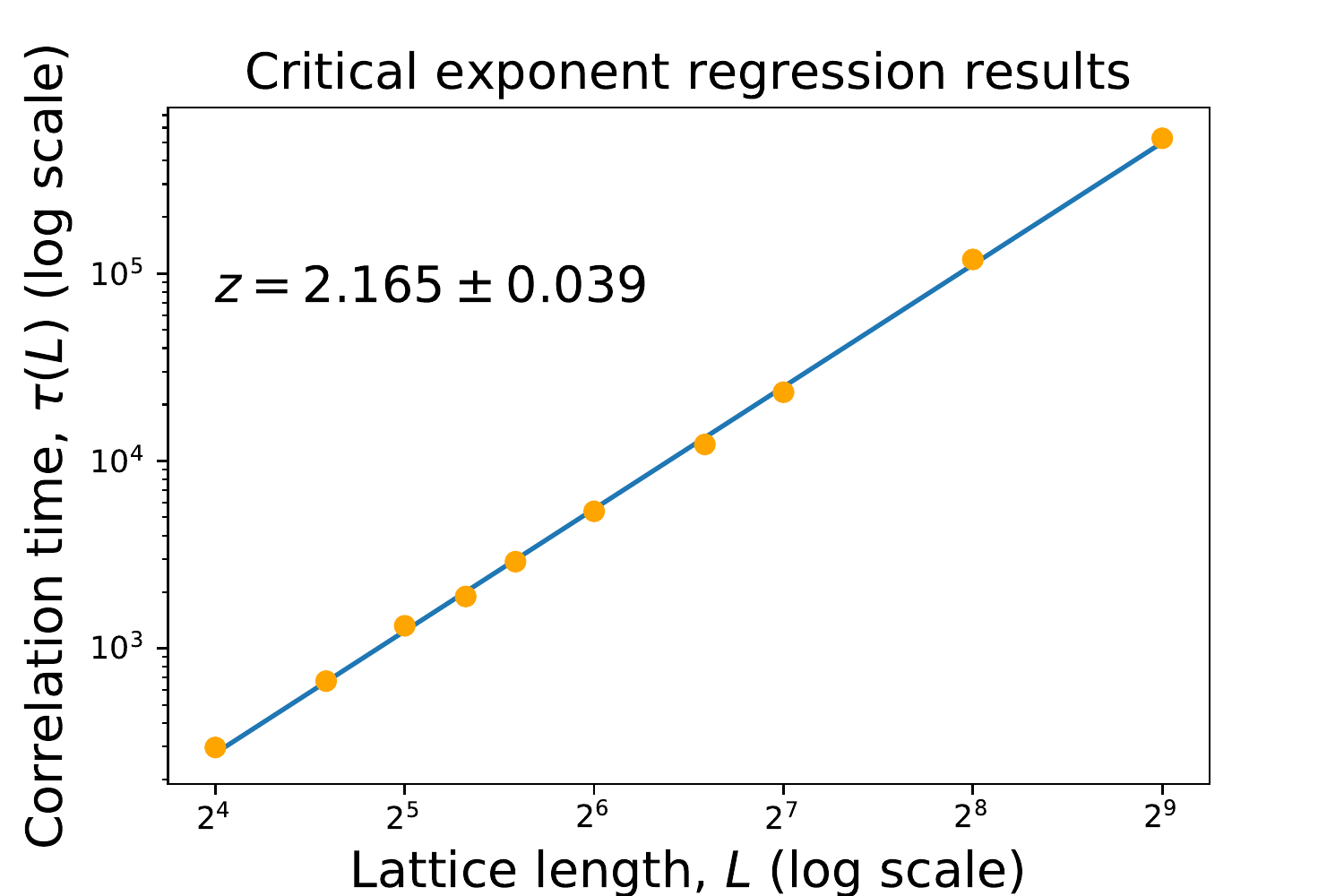}
 \caption{{\it Critical exponent calculation with a Quantum Random Numbers source.} Orange: estimated relaxation times (in log10 scale) as a function of the lattice size (in log2 scale); blue, the linear fitting $\log_2(\tau) = z\cdot \log_2(L) + b$ associated. The slope of the fitting $z=2.165\pm0.039$ corresponds to the critical exponent, in agreement with theoretical results.}
  \label{fig:fig8}
\end{figure}

In this section, we present the results on the calculation of the dynamical critical exponent using a FPGA and, most importantly, a quantum RNG (QRNG). As in previous sections, we performed simulations for different lattice sizes. In Fig.~\ref{fig:fig7} we show the  time-delayed correlation of the order parameter (magnetization),  $\chi(t)$, as a function of time. We see the expected exponential decay described by Eq.~\eqref{eq:chi}. By fitting the previous equation, we extract the average value of the exponent $\tau$ for each $L$. We performed 100 simulations for $L$ up to 256 and  15 simulations for $L=512$ due to the large computational cost in this last case (see discussion below). It is very apparent from last panel in Fig.~\ref{fig:fig7} that the results for $L=512$ are more noisy that those obtained for smaller lattice sizes, probably due to the smaller number of simulations.

In Fig.~\ref{fig:fig8} we represent the average correlation time $\tau(L)$ obtained from these results. Performing a linear fitting of the obtained curve, we  find a value of the critical exponent $z=2.165\pm0.039$. We emphasize that, from here, there is a clear strategy to improve this value: performing multiple simulations for bigger lattices would add more points to the fit, resulting in a better approximation of this coefficient. However, unlike the case of pseudo-generators, obtaining correlation times for larger cell sizes is limited by the vast need for random numbers required by the simulation. Note that, for each step of the simulation, $N \propto L^2$ random numbers are required. As the simulations are run for $1300\tau_k = 1300 L^z$ steps, we require on the order of $1300 L^{z+2}$ random numbers. 
Assuming that each of these numbers has 32-bit precision, we face a massive consumption of about $41600\cdot L^{z+2}$ randomly-distributed bits. For small cell sizes, these requirements are innocuous; however, the $\sim L^4$ exponential growth in demand for random numbers is prohibitive. In the case of Quside's QRNG apparatus used in our simulations, which reaches quantum random number generation rates of 400 Mbps, we spend around 12 hours for each simulation of $L=256$. Doubling the size of the network ($L=$512) requires twenty days for each simulation; by doubling it again ($L=1024$), we estimate a simulation time of almost six months per simulation. To avoid this computational bottleneck, and speeding-up the simulation of the $L=512$ case, we used an amplification of the QRNG's random numbers. This amplification consists on the implementation of a PRNG on the FPGA, which is reseeded as fast as the QRNG provides new seeds. . This decision introduces some correlations that are not present for smaller lattices. This fact, along the very low number of repetitions, could potentially be the reason of the more noisy behaviour observed in the last panel of fig.\ref{fig:fig7}, as discussed in the PRNG section. Nevertheless, due to the small size of the statistical sample at hand, we cannot conclude which one of them is the predominant reason without additional simulations.

\section{Values of the dynamical critical exponent in the scientific literature}

\begin{figure}[ht]
    \centering
    \includegraphics[width=\columnwidth]{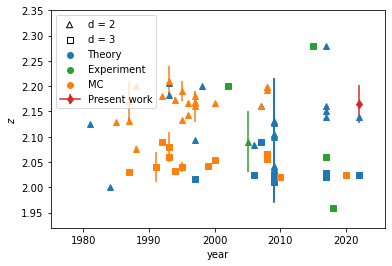}\\
    \includegraphics[width=\columnwidth]{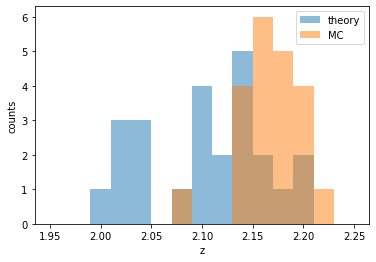}
    \caption{{\it Dynamical critical exponents given in the literature}. Top: Different $z$ calculated over the years in two and three dimensions. Triangles and squares label two- and three- dimensional lattices, while colors blue, green and orange mark whether these values were obtained theoretically, experimentally or via MC simulations, respectively. The value obtained in the present work is marked with a red diamond. Bottom: Histogram showing the $z$ predictions theoretically and with MC in two dimensions.  }
    \label{fig:fig9}
\end{figure}

Before concluding, we offer in this section a  discussion and review of previous results. 
Over the years, many attempts to give an appropriate value for the dynamical critical exponent $z$ have been carried on from theoretical, experimental and MC sides. Here, with the aim of illustrating how vastly the obtained results vary, we present a long, yet non-exhaustive collection of values found in the literature, for both two- and three- dimensional lattices. Some of the references presented here give various values corresponding to different types of lattices, in an attempt to show the postulated universality of $z$ across models. We plot in Fig.~\ref{fig:fig9} the obtained exponents as a function of the year of its calculation. Surprisingly, even with the expected improvement in the used methods, there is no clear tendency, neither in two-, three-dimension, theoretical or MC calculations, and even the various results obtained in recent years show a wide spreading. The data is gathered in Tables~\ref{tab:tab6} and \ref{tab:tab7} from Appendix~\ref{sec:data}.  We plot in Fig.~\ref{fig:fig9}b a histogram grouping the theoretical predictions and the MC ones. While the values obtained by means of theoretical methods do not show any apparent distribution, the ones obtained with MC calculations can be fitted to a Gaussian distribution with mean $\langle z_{\rm{MC}}\rangle=2.1664$. Importantly, such value is close to the one predicted by our QRNG calculations. While this is not conclusive, it clearly shows the importance of pushing forward the research conducted in this paper to larger lattice sizes and more repetitions, given that the QRNG do not have the variance limitation of the PRNG. 

\section{Conclusions and outlook}

In this work, we have studied the effect of different random number sources in highly complex computational tasks. We have shown that the correlations appearing in pseudo-random number sequences greatly affect the results obtained when computing the dynamic critical exponent of the Ising model. First, we showcased that once the period of repetition of the PRNG is reached (and therefore significant correlations start to appear), the variance of our results explodes. Nonetheless, we have also shown that, reducing the correlations introduced by the repetition of the sequences by reseeding our generators, such variance explosion can be avoided, reaching a value that depends on the frequency at which the reseeding is carried on. In this direction, by extrapolating the values obtained for different reseeding frequencies, we have also estimated an upper bound for the variance that should be expected when using a true random number generator.
The reseeding of the generators also affects the mean value of the obtained dynamic exponent, making it closer to the theoretical estimate. We can therefore state that, the closer a RNG is to a true RNG, the better the estimation of the dynamics exponent, both in terms of mean and variance. 

To avoid the aforementioned problems, we have shown a new promising avenue: the use of quantum random number generator. The first explorations on the matter, presented in this work, yield a mean value of the critical exponent $z$ in accordance with majority of results proposed in literature. While the results obtained in this work are encouraging and show the potential of QRNG for complex simulations, we believe that further improvements need to be done in order to be conclusive on this value. For this, the random number generation speed is still the current simulation bottleneck, as shown in Appendix~\ref{sec:numberofP}, making it difficult to get enough statistics when using properly guaranteed, quantum randomness. Nonetheless, the recent advances in the field point towards such achievement. Then, if it is shown that the QRNG results converge towards a stable and conclusive critical exponent for large lattices,  yielding a variance equal or lower to the bound estimated in this work, the use of the Ising model's critical exponent calculation as a benchmarking tool for other randomness generators will be greatly justified, by merely contrasting the latter with the results obtained by the QRNG.

%Further details on the implementation and results associated with PRNG discrimination can be found in the appendices. Finally, when comparing such results with the ones obtained using random numbers of quantum origin, we observe that BLABLA, fig. X. \MAGM{Should talk about this later? }

%%%%%%%%%%%%%%%%%%%%%%%%%%%%%%%%%%%%%%%
\section*{Acknowledgments}
%%%%%%%%%%%%%%%%%%%%%%%%%%%%%%%%%%%%%%%

We acknowledge Joana Frexanet, Lluis Torner, Sergi Ferrando, Pau Gómez, Felix Tebbenjohanns, and Carlos Abellán.
We acknowledge also discussions with Josep Maria Martorell, Mervi Mantsinen, Xavier Saez, Vassil Alexandrov, Francisco Castejón, and Shinsuke Satake in the early stage of this project.
We acknowledge 
%the Spanish Ministry MINECO (National Plan 15 Grant: FISICATEAMO No. FIS2016-79508-P, SEVERO OCHOA No. SEV-2015-0522, FPI), European Social Fund, Fundaci\'o Cellex, Generalitat de Catalunya (AGAUR Grant No. 2017 SGR 1341 and CERCA/Program), ERC AdG NOQIA, and the National Science Centre, Poland-Symfonia Grant No. 2016/20/W/ST4/00314. 
ERC AdG NOQIA; Agencia Estatal de Investigaci\'on (R\&D project CEX2019-000910-S, funded by MCIN/ AEI/10.13039/501100011033, Plan Nacional FIDEUA PID2019-106901GB-I00, FPI, QUANTERA MAQS PCI2019-111828-2, Proyectos de I+D+I ''Retos Colaboraci\'on'' QUSPIN RTC2019-007196-7), MCIN via European Union NextGenerationEU (PRTR-C17.I1);  Fundaci\'o Cellex; Fundaci\'o Mir-Puig; Generalitat de Catalunya through the European Social Fund FEDER and CERCA program (AGAUR Grant No. 2017 SGR 134, QuantumCAT / U16-011424, co-funded by ERDF Operational Program of Catalonia 2014-2020); EU Horizon 2020 FET-OPEN OPTOlogic (Grant No 899794); National Science Centre, Poland (Symfonia Grant No. 2016/20/W/ST4/00314); European Union's Horizon 2020 research and innovation programme under the Marie-Sk\l{}odowska-Curie grant agreement No 101029393 (STREDCH) and No 847648  (''La Caixa'' Junior Leaders fellowships ID100010434: LCF/BQ/PI19/11690013, LCF/BQ/PI20/11760031, LCF/BQ/PR20/11770012, LCF/BQ/PR21/11840013).
D.C.G. acknowledges funding from Generalitat de Catalunya (AGAUR Doctorats Industrials 2019, 2n termini). 
MAGM acknowledges funding from the Spanish Ministry of Education and Professional Training (MEFP) through the Beatriz Galindo program 2018 (BEAGAL18/00203). G.M-G. acknowledges support from the Austrian Science Fund (FWF) through SFB BeyondC F7102.

\appendix

\section*{Appendices}

\section{Parallelization using FPGAs/GPUs: required number of random numbers}
\label{sec:numberofP}

Note that our code runs over every single spin at any time step. This updating process differs from other possible implementations of the Metropolis algorithm that rely on sequential updates such as sampling and updating $L^2$ (with $L$ the lattice size) spins for each iteration, with the spins selected at random. Such a sequential implementation does not guarantee that all of the spins will be given the chance of being updated, nor that all of them will be updated only once in each Monte Carlo sweep (even if it does satisfy ergodicity in the long run). This fact does not make a difference in the obtained results \cite{Weigel2012}, but yields a considerable improvement in performance when using the parallel Metropolis version. Not only it allows the spins to be updated in a parallel manner, but it also reduces the entropy consumption of the algorithm since it does not require such $L^{2}$ random numbers at every time step that chooses the spins to be updated. On the other hand, our implementation assumes that every single spin might be asked to flip with a certain probability following the Metropolis algorithm, thus consuming again $L^{2}$ random numbers, while a CPU-based code would only need a fraction $\alpha \le 1$ of them, corresponding to these spins that are not flipped directly after computing its energy difference between both initial and proposed states.

Taking everything into account, a sequential algorithm consumes $n_{\rm{seq}} = L^{2}\left(1+\alpha\right)$ random numbers at every time step, while the parallel one only demands $n_{\rm{par}} = L^{2}$, which equals the best lower bound achievable with a sequential device. In this direction, our code could be further optimized in order to avoid the discard of the $L^{2}\left(1-\alpha\right)$ numbers that are not used in the previous iteration, replacing these ones consumed, and therefore yielding a much lower demand of $n_{\rm{par}} = L^{2}\alpha$.

\begin{table}[ht]
\begin{tabular}{|c|c|c|c|c|c|}
\hline
\tiny\textbf{L} &
  \textbf{\tiny \begin{tabular}[c]{@{}l@{}}MC updates/\\ simulation\end{tabular}} &
  \textbf{\tiny $N_{it}$} &
  \textbf{\tiny \begin{tabular}[c]{@{}l@{}}Random bits/\\ update\end{tabular}} &
  \textbf{\tiny \begin{tabular}[c]{@{}l@{}}Total bits\\ required (GB)\end{tabular}} &
  \textbf{\tiny \begin{tabular}[c]{@{}l@{}}Simulation time \\ (days)\end{tabular}} \\ \hline
16  & 533192    & 100 & 512            & 3.2      & 0.001 \\ \hline
24  & 1285292   & 100 & 1152           & 17.2     & 0.006 \\ \hline
32  & 2399489   & 100 & 2048           & 57.2     & 0.02  \\ \hline
48  & 5784114   & 100 & 4608           & 310.3    & 0.1   \\ \hline
64  & 10798263  & 100 & 8192           & 1029.8   & 0.3   \\ \hline
96  & 26029866  & 100 & 18432          & 5585.4   & 1.9   \\ \hline
128 & 48594709  & 100 & 32768          & 18537.4  & 6.1   \\ \hline
256 & 218687559 & 100 & 131072         & 333690.7 & 111   \\ \hline
512 & 984145175 & 10  & 524288         & 600674.5 & 199   \\ \hline
    &           &     & \textbf{TOTAL} & 959905.8 & 318   \\ \hline
\end{tabular}
\caption{Estimated randomness consumption and simulation times for the QRNG results.}
\label{tab:sim_times}
\end{table}

%\section{ Pseudo-randomness: GPU implementation.}
%\label{sec:PRNG_GPU}
When implementing the code in a GPU, we note that,  given the availability of GPU resources, the possible optimization of the PRN generation introduced in previous paragraph would not yield any improvement. It is also worth noting that such optimization would introduce even more correlations in the results: by possibly discarding some random numbers, we indeed lower the correlation between the probabilities used by a given spin to be updated at two different times.

\section{Linear fit of the dynamic exponent.}
As discussed in previous sections, the relaxation time of the system is expected to have a power-law relation with the lattice size $L$. In order to obtain the dynamic exponent $z$, one can take logarithms to both sides of Eq.~\eqref{eq:tauFSS} by naively reducing it to an equality $\tau = L^z$ and then make a linear fit ${\rm log}(\tau) = z {\rm log}(L)$. Instead, we strictly consider the proportional sign in Eq.~\eqref{eq:tauFSS} by stating that $\tau = \tau_0 L^z$, therefore allowing the linear fit to have an offset, ${\rm log}(\tau) = z {\rm log}(L) + {\rm log}(\tau_0)$.

We summarize our findings in Section~\ref{tab: log(tau0)}, in which we compare the offset obtained by the fitting of the QRNG data against those obtained by the PRNG data. There are two main things worth noting about the obtained results. First, we observe that all of them yield a non-zero, negative offset. And, secondly, we note that those PRNG using reseeding (thus having an infinite period, and therefore being closer to a TRNG) yield values closer to the one obtained by the QRNG. This fact hints that this parameter could potentially serve in the purpose of discerning good from bad randomness too.

\begin{table}[ht]
    \centering
    \begin{tabular}{|c|c|c|}
         \hline
         & $log(\tau_0)$\\
         \hline
         QRNG & $-0.362$\\
         \hline
         PRNG3 $k=\tfrac{1}{2}$ & $-0.1701$\\
         \hline
         PRNG3 $k=2$ & $-0.2454$\\
         \hline
         PRNG2 & $-0.0906$\\
         \hline
         PRNG0 & $-0.1007$\\
         \hline
         PRNG1 & $-0.1183$\\ 
         \hline
    \end{tabular}
    \caption{Results obtained for $log(\tau_0)$ by using different RNG. These values correspond to the fittings yielding the dynamic exponents shown in previous sections.}
    \label{tab: log(tau0)}
\end{table}

\section{Summary of all calculated dynamical critical exponents, for two and three dimension.}
\label{sec:data}

In Tables~\ref{tab:tab6} and \ref{tab:tab7} we gather all calculated exponents from the literature in two and three dimensions, to the best of our knowledge, for theoretical, MC as well as experimental approaches. This is the data plotted in Fig.~\ref{fig:fig8}.  
\begin{table}[ht]
    \centering
    \begin{tabular}{|c|c|c|c|} 
        \hline
         \textbf{Year} & \textbf{Reference} & \textbf{Method} & \textbf{$z$}  \\
         \hline
         1981 & Bausch & Theory & $2.126$ \\
         \hline
         1984 & Domany & Theory & $2$ \\
         \hline
        1985 & Williams & MC & $2.13(3)$ \\
        \hline
        1987 & Ito & MC & $2.132\pm0.008$ \\
        \hline
        1987 & Tang & MC & $2.17\pm0.04$ \\
        \hline
        1988 & Ito & MC & $2.2$ \\
        \hline
        1988 & Mori & MC & $2.076\pm0.005$ \\
        \hline
        1992 & Stauffer & MC & $2.18$ \\
        \hline
        1993 & Dammann & Theory & $2.183\pm0.005$ \\
        \hline
        1993 & Wang & Theory & $2.207\pm0.008$ \\
        \hline
        1993 & Muenkel & MC & $2.21\pm0.03$ \\
        \hline
        1994 & Grassberger & MC & $2.172\pm0.006$ \\
        \hline
        1995 & Gropengiesser & MC & $2.18\pm0.02$ \\
        \hline
        1995 & Li & MC & $2.1337(41)$ \\
        \hline
        1996 & Li & MC & $2.143(5)$ \\
        \hline
        1996 & Nightingale & MC & $2.1665(12)$ \\
        \hline
        1997 & Soares & MC & $2.16\pm0.03$ \\
        \hline
        1997 & Wang & MC & $2.168\pm0.005$ \\
        \hline 
        1997 & Wang & MC & $2.180\pm0.009$, TP \\
        \hline
        1997 & Wang & MC & $2.167\pm0.008$, hc \\
        \hline
        1997 & Prudnikov & Theory & $2.093$ \\
        \hline
        1998 & Wang & Theory & $2.2$ \\
        \hline
        2000 & Nightingale & MC & $2.1667\pm0.0005$ \\
        \hline
        2005 & Dunlavy & Experiment & $2.09\pm0.06$ \\
        \hline
        2006 & Krinitsyin & Theory & $2.0842\pm0.0039$ \\
        \hline
        2007 & Canet & Theory & $2.16(1)$ \\
        \hline
        2007 & Lei & MC & $2.16$ \\
        \hline
        2008 & Murase & MC & $2.193(5)$ \\
        \hline
        2008 & Murase & MC & $2.198(4)$, hc \\
        \hline
        2008 & Murase & MC & $2.199(3)$, TP \\
        \hline
        2009 & Nalimov & Theory & $2.020\pm0.045$ \\
        \hline
        2009 & Nalimov & Theory & $2.023\pm0.053$ \\
        \hline
        2009 & Nalimov & Theory & $2.026\pm0.055$ \\
        \hline
        2009 & Nalimov & Theory & $2.100\pm0.089$ \\
        \hline
        2009 & Nalimov & Theory & $2.105\pm0.084$ \\
        \hline
        2009 & Nalimov & Theory & $2.104\pm0.080$ \\
        \hline
        2009 & Nalimov & Theory & $2.127\pm0.089$ \\
        \hline
        2009 & Nalimov & Theory & $2.132\pm0.084$ \\
        \hline
        2009 & Nalimov & Theory & $2.130\pm0.080$ \\
        \hline
        2009 & Nalimov & Theory & $2.037^{+0.033}_{-0.0}$ \\
        \hline
        2009 & Nalimov & Theory & $2.041^{+0.040}_{-0.0}$ \\
        \hline
        2009 & Nalimov & Theory & $2.042^{+0.041}_{-0.0}$ \\
        \hline
        2017 & Duclut & Theory & $2.28$  \\
        \hline
        2017 & Duclut & Theory & $2.16$ \\
        \hline
        2017 & Duclut & Theory & $2.15$ \\
        \hline
        2017 & Duclut & Theory & $2.14$ \\
        \hline
        2022 & Adzhemyan & Theory & $2.14(2)$ \\
        \hline
        
    \end{tabular}
    \caption{All calculated exponents from the literature in  two dimensions, to the best of our knowledge.}
    \label{tab:tab6}
\end{table}

\begin{table}[ht]
    %\centering
    \begin{tabular}{|c|c|c|c|} 
        \hline
         \textbf{Year} & \textbf{Reference} & \textbf{Method} & \textbf{$z$}  \\
         \hline
         1987 & Wansleben & MC & $2.03\pm0.04$ \\
         \hline
         1991 & Wansleben & MC & $2.04\pm0.03$ \\
         \hline
         1992 & Stauffer & MC & $2.09$ \\
         \hline
         1993 & Ito & MC & $2.06(2)$ \\
         \hline
         1993 & Muenkel & MC & $2.08\pm0.03$ \\
         \hline
         1994 & Grassberger & MC & $2.032\pm0.004$ \\
         \hline
         1995 & Gropengiesser & MC & $2.04\pm0.01$ \\
         \hline
         1997 & Prudnikov & Theory & $2.017$ \\
         \hline
         1999 & Jaster & MC & $2.042(6)$ \\
         \hline
         2000 & Ito & MC & $2.055(10)$ \\
         \hline
         2002 & Livet & Experiment & $2.2$ \\
         \hline
         2006 & Krinitsyin & Theory & $2.0237\pm0.0055$ \\
         \hline
         2007 & Canet & Theory & $2.09(4)$ \\
         \hline
         2008 & Murase & MC & $2.065(25)$, bcc \\
         \hline
         2008 & Murase & MC & $2.057(25)$, fcc \\
         \hline
         2009 & Nalimov & Theory & $2.011\pm0.012$ \\
         \hline
         2009 & Nalimov & Theory & $2.013\pm0.012$ \\
         \hline
         2009 & Nalimov & Theory & $2.014\pm0.011$ \\
         \hline
         2009 & Nalimov & Theory & $2.021\pm0.006$ \\
         \hline
         2009 & Nalimov & Theory & $2.022\pm0.005$ \\
         \hline
         2009 & Nalimov & Theory & $2.022\pm0.005$ \\
         \hline
         2009 & Nalimov & Theory & $2.023\pm0.006$ \\
         \hline
         2009 & Nalimov & Theory & $2.024\pm0.005$ \\
         \hline
         2009 & Nalimov & Theory & $2.024\pm0.005$ \\
         \hline
         2009 & Nalimov & Theory & $2.013^{+0.011}_{-0.0}$ \\
         \hline
         2009 & Nalimov & Theory & $2.014^{+0.011}_{-0.0}$ \\
         \hline
         2009 & Nalimov & Theory & $2.014^{+0.011}_{-0.0}$ \\
         \hline
         2010 & Collura & MC & $2.020(8)$ \\
         \hline
         2015 & Livet & Experiment & $2.28$ \\
         \hline
         2017 & Niermann & Experiment & $2.06$ \\
         \hline
         2017 & Duclut & Theory & $2.029$ \\
         \hline
         2017 & Duclut & Theory & $2.024$ \\
         \hline
         2017 & Duclut & Theory & $2.023$ \\
         \hline
         2017 & Duclut & Theory & $2.025$ \\
         \hline
         2017 & Duclut & Theory & $2.021$ \\
         \hline
         2017 & Duclut & Theory & $2.021$ \\
         \hline
         2018 & Livet & Experiment & $1.96(11)$ \\
         \hline
         2020 & Hasenbusch & MC & $2.0245(15)$ \\
         \hline
         2022 & Adzheyman & Theory & $2.0235(8)$ \\
         \hline
         
    \end{tabular}
    \caption{All calculated exponents from the literature in three dimensions, to the best of our knowledge.}
    \label{tab:tab7}
\end{table}

\bibliographystyle{apsrev4-2}
\bibliography{biblio}

\end{document}